\documentclass[twocolumn]{aastex63}
\usepackage{amsmath}
\usepackage{graphicx}
\usepackage{mathtools}
\usepackage{tabularx}
\usepackage{halloweenmath}

\graphicspath{{./}{figures/}}

\begin{document}
\correspondingauthor{Mariannne Cowherd}
\email{cowherd@berkeley.edu}

\title{The Atmospheric Ghost Correction}

\author[0000-0002-3165-4504]{Mariannne Cowherd}
\affil{Department of Environmental Science, Policy, and Management, University of California\textendash Berkeley, Berkeley, CA, 94720, USA}
\author[0000-0003-0217-3880]{Isaaac Malsky}
\affil{Department of Astronomy and Astrophysics, University of Michigan, Ann Arbor, MI, 48109, USA}

\begin{abstract}

While the presence of ghosts has been known for decades, the impact of these apparitions on remote sensing observations has gone unquantified, leaving atmospheric corrections susceptible to ghosting. In this work, we present the first spectral characterization of three common ghost types and provide a framework for incorporating these properties into atmospheric correction algorithms. We demonstrate the effect of this improved atmospheric ghost correction compared to an atmosphere-only implementation. Lastly, we discuss some preliminary spatiotemporal variations in haunting intensity and type.

\end{abstract}


\section{Introduction} \label{sec:intro}

Satellite remote sensing of the Earth surface allows us to observe land and water from above semi-continuously in time and space. Technology both propels and constrains us; we record data from all sources on the Earth but must also account for undesired data sources\footnote{Common examples of undesired data include elevation reported in units of survey feet and suggestions to ``just use machine learning it's easy''}. For surface processes, the presence of the atmosphere, while technically necessary for life, burdens observed spectra with absorption and scattering features from clouds, haze, and aerosols. Atmospheric correction is therefore a necessary step in producing usable remotely sensed products. Significant development of atmospheric correction algorithms has occurred since at least the beginning of satellite remote sensing by humans \citep{saastamoinen1972atmospheric}, if not since the advent of time itself$^\text{[citation needed]}$.

In general, an atmospheric correction algorithm is based on the premise that the radiance received at a sensor is related but not identical to the radiance of the surface:
\begin{equation}
L_s = H \rho T + L_p
\end{equation}
where $L_s$ is the radiance at the sensor, $H$ is incoming radiance from the Sun, $\rho$ is the surface reflectance, $T$ is the transmission through the atmosphere, and $L_p$ is the radiance of the atmosphere along the sensor's path.

Realistically, we must estimate many of these components based on truly observable quantities and so the applied processing steps are complex. Common methods, many of which can be used sequentially if you can't pick, include dark object subtraction \citep{chavez1988improved}, (pseudo)-invariant features calibration \citep{cosnefroy1996selection}, radiometric control sets \citep{hall1991radiometric}, and modeling based on atmospheric conditions and radiative transfer \citep{berk1999modtran4}, among others. However, all published methods to date neglect to account for the unique spectral properties of ghosts above the Earth surface. Atmospheric specters can dominate surface signals in some wavelengths, which has the effect of masking the desired message, i.e. ghosting \citep{ex2016}.

Ghosts, which are apparitions of dead organisms, consist of manifest energy. While it is likely that this energy spans a significant portion of the electromagnetic spectrum, we focus on only the visible through shortwave infrared range in this work. It is also notable that ghosts contain psychokinetic energy (PKE), and while this is immensely important for the detection and trapping of ghosts \citep{parker2018you}, PKE is not suspected to impact Earth system remote sensing of electromagnetic radiation.

In this work, we provide the first characteristic spectra of three ghostly archetypes for the first time and provide guidance for atmospheric ghost corrections. We also analyze spatiotemporal patterns in ghosts to inform future interpretation of remote sensing data.

\begin{figure}
    \centering
    \includegraphics[height = 8cm, width=\linewidth,trim={0 0 9cm 0},clip]{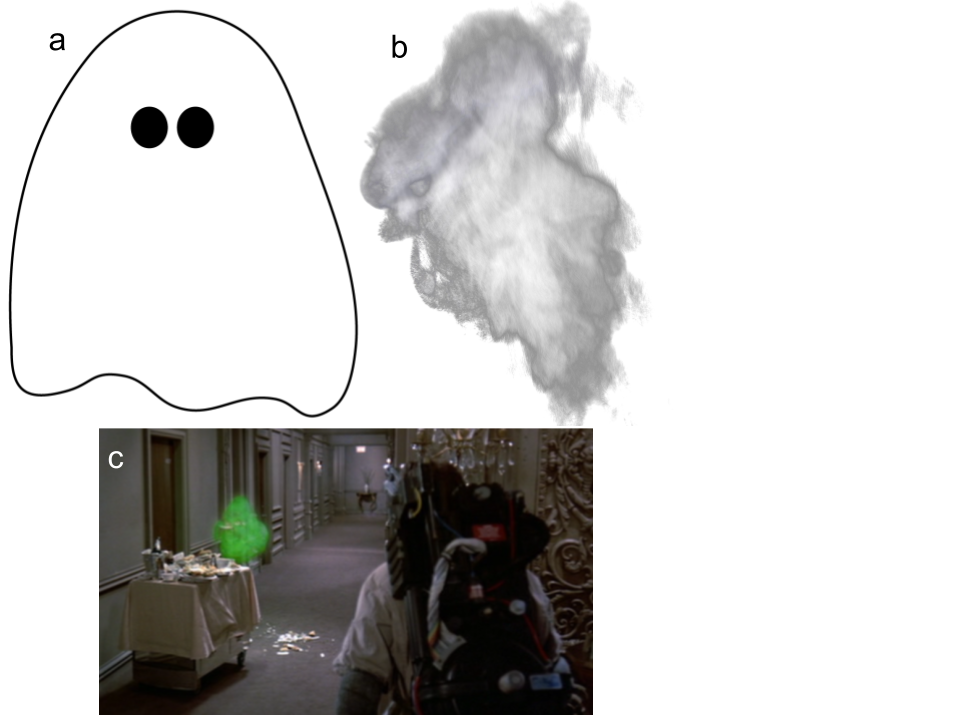}
    \caption{Images of the three ghost archetypes: highly visible ghosts (a), wisps of energy (b), and ectoplasmic green ghosts (c). Images were taken \textit{in situ} and post-processed in ENVI with the Valencia filter.}
    \label{fig:ghosts}
\end{figure}

\section{Methods} \label{sec:methods}
To characterize specter spectra, we conducted laboratory spectrometry on ghost specimens. We captured ghosts in the wild using standard methods from \cite{ghostbusters1984} and refined in \cite{ghostbusters2016}. No IRB approval was necessary. We then measured samples using a field spectroradiometer. After imaging, ghosts were given a snack and released into the wild.
\begin{equation}
    NGI = \frac{NIR-\mathghost}{NIR+\mathghost}
\end{equation}
 Supernaturally, we extend this to define difference normalized ghost index (dNGI) and relative difference normalized ghost index (RdNGI) as comparisons of pre- and post- event spookiness. While we use NGI to observe the usual haunts, dNGI or RdNGI are appropriate for evaluating the impact of a specific event (e.g. a s\'eance or very windy night).

\section{Results} \label{sec:results}
\begin{figure}
    \centering
    \includegraphics[width=\linewidth]{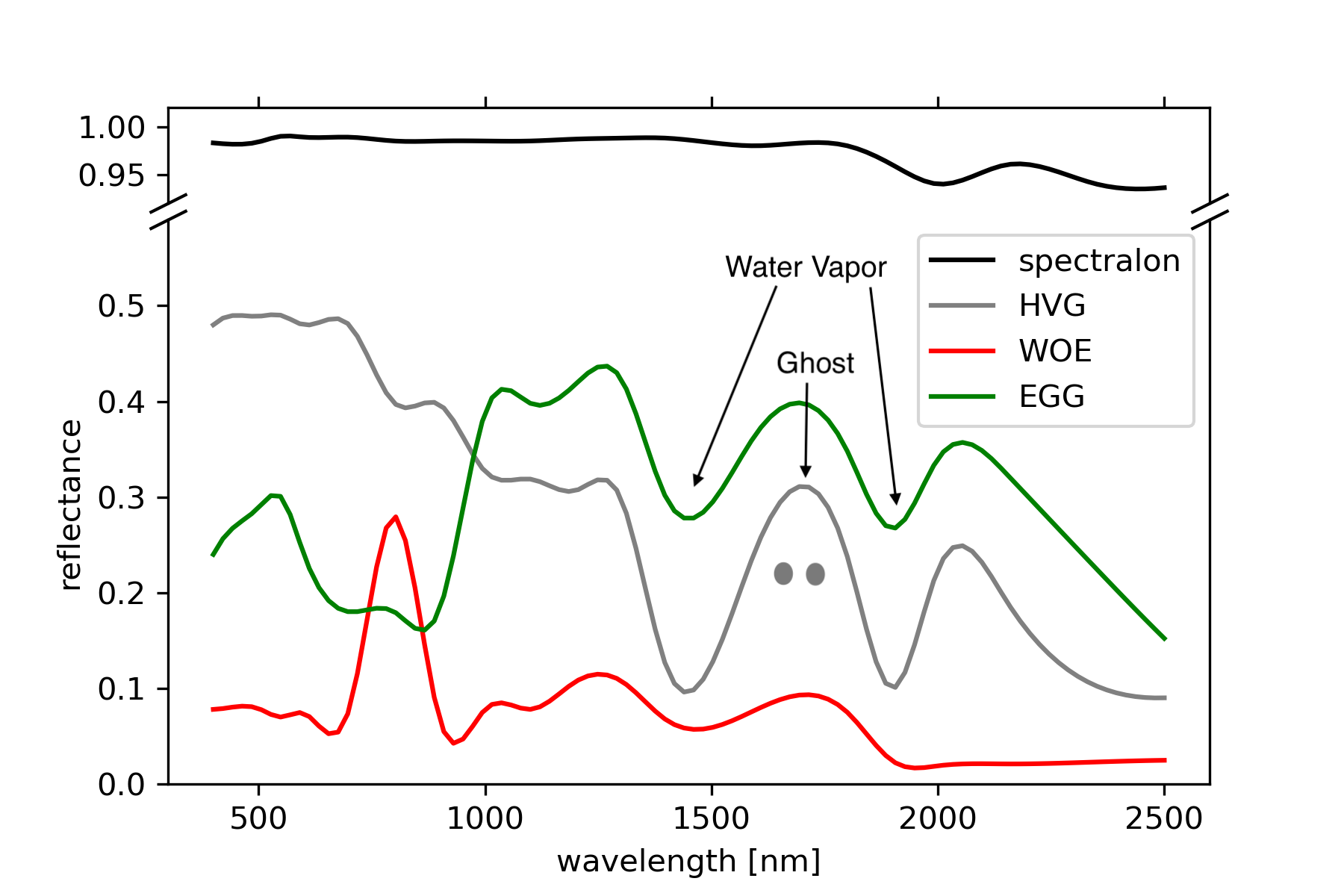}
    \caption{Specter signatures of highly visible ghosts (gray), wisps of energy 
    (red), and ectoplasmic green ghosts (green) with spectralon (black) and an oak leaf (invisible) for reference.}
    \label{fig:spectra}
\end{figure}

The true spectral signature of ghosts varies with the composition of the individual ghost. As this remote sensing is ``multi-spectral'' we must consider the impact of more than one specter. Here, we characterize three archetypes: Highly Visible Ghosts (HVGs),  and Wisps Of Energy (WOEs), and Ectoplasmic Green Ghosts (EGGs).
Figure \ref{fig:spectra} shows representative spectra for all three ghost archetypes, with a spectralon reading for reference. We see relatively high reflectance in the visible range for HVGs, as expected from their white exterior (except for the dark ovals shown in Figure \ref{fig:ghosts}a, but these will only appear in sideways-acquired signals). EGGs show distinctive reflectance in the green portion of the visible range (500-570 nm). Both HVGs and EGGs show water absorption features around 1400 and 1900 nm. This is expected from EGGs, as they are made of slime \cite{ono1978ghosts}, but has not been previously observed from HVGs. This is discussed further in Section \ref{sec:concl}.

We propose an updated ghost-atmospheric correction
\begin{equation}
    L_{s,\mathghost} = H \rho T_aT_{\mathghost} + L_p L_{\mathghost}
\end{equation}
where  $T_{\mathghost}$ is the transmission through ghostly forms, calculated generally from
\begin{equation}
   T_{\mathghost} = \int_{0}^{\mathbat} \Big[\frac{1}{2}\mathcloud({\mathrightghost + \mathleftghost}) \Big] d\mathghost
\end{equation}
with the shapes of $\mathleftghost$ and $\mathrightghost$ dependent on ghost archetype and concentration and $\mathcloud$ conceals a fudge factor.

The path-line ghost radiance $L_{\mathghost}$ is calculated
\begin{equation}
   L_{\mathghost}(\lambda) = \mathwitch \Big\langle \mathghost^2(\lambda)\Big\rangle
\end{equation}
for wavelengths $\lambda$. The witch function is encoded in Fortran and included in the supplement\footnote{It is not.}.

\begin{figure}
    \centering
    \includegraphics[width=\linewidth]{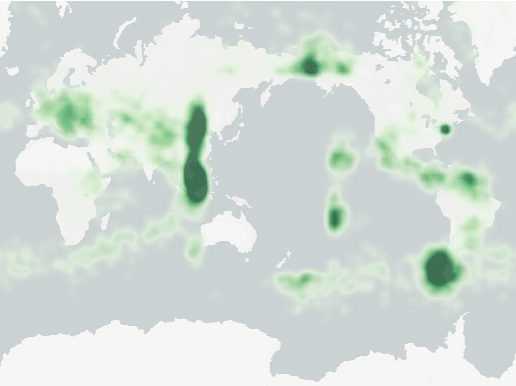}
    \caption{Map of Normalized Ghost Index (NGI) averaged for water year 2016.}
    \label{fig:map}
\end{figure}
Figure \ref{fig:map} shows spatial variations in NGI for water year 2016. Notably, we see a hot spot in New York City, the so-called ghost capital of the world. We clearly observe the specter haunting Europe. There are no ghosts over Antarctica; they may have escaped to space through the ozone hole or they may prefer warmer weather.

Spatial distribution and overall apparition magnitude varies temporally. Seasonal variance dominates inter-annual variance globally (data lost). Ghosts peak in late October and early November for unknown reasons.

\section{Discussion} \label{sec:disc}
Spectral signatures of ghosts\footnote{Boo!} from laboratory work indicate their potential to contribute to radiance signals received by the satellites orbiting the Earth. We show that ghostly overlap with wavelength ranges studied used for Earth surface research and confirm the presence of water lines in EGGs. The detection of water absorption lines for HVGs suggests that there may be something underneath the sheet, but we did not think to check this before releasing the ghost.

Volcanic activity, the Bering Land Bridge, and circumpolar circulation are implicated as sources of ghosts due to spatial correlation. High winds likely blew all ghosts from the Sahara region to the west. If the flow of ghosts is like the flow of air, then this would be an easterly ghost-flow. However, if the flow of ghosts is like the flow of water, then this would be a westerly ghost-flow. We recommend against further discourse in the community to standardize terminology.

The spatial variation in ghosts is highly time-variable. We cannot definitively say that a given location is haunted. Rather, we encourage NGI as a proxy for where the ghostly terms ($T_{\mathghost}$ and $L_{\mathghost}$) are likely significant in Equation 3.

\section{Conclusion} \label{sec:concl}
In this paper we explore the impact of ghosts on atmospheric correction in multispectral remote sensing and demonstrate the need for an atmospheric ghost correction. The impact of ghosts on radiance observations in the haunted wavelength range is semi-clearly present in many regions of the world and is heterogeneous in space and time. This should not necessarily motivate or require further research.

\section*{acknowledgments}
The authors deny any allegations of funding.

\bibliographystyle{aasjournal}
\bibliography{main}

\end{document}